\documentclass[preprint]{revtex4}
\usepackage[utf8]{inputenc}
\usepackage{natbib}
\usepackage{amssymb}
\usepackage[english]{babel}
\usepackage{lineno,hyperref}
\usepackage[utf8]{inputenc}
\usepackage{amsmath}
\usepackage{graphicx}
\usepackage[colorinlistoftodos]{todonotes}
\usepackage{rotating}
\usepackage{color}
\definecolor{red}{rgb}{1,0,0}
\definecolor{blue}{rgb}{0,0,1}
\usepackage{pdflscape}
\usepackage{rotating}
\usepackage{ulem}
\usepackage{algorithm}
\usepackage{algpseudocode}

\begin{document}

\title{Fast Switch and Spline Scheme for Accurate Inversion of Nonlinear Functions:  \\
The New First Choice Solution to Kepler's Equation}

\author{Daniele Tommasini$^1$ and David N. Olivieri$^2$\\
\small $^1$Applied Physics Department, School of Aeronautic and Space Engineering, Universidade de Vigo. As Lagoas s/n, Ourense, 32004 Spain.\\
$^{2}$Computer Science Department, School of Informatics (ESEI), Universidade de Vigo. As Lagoas s/n, Ourense, 32004 Spain\\\
daniele@uvigo.es; olivieri@uvigo.es}

\date{\today}


\begin{abstract}
Numerically obtaining the inverse of a function is a common task for many scientific problems, often solved using a Newton iteration method.  Here we describe an alternative scheme,  based on switching variables followed by spline interpolation, which can be applied to monotonic functions under very general conditions. To optimize the algorithm, we designed a specific ultra-fast spline routine. We also derive analytically the theoretical errors of the method and test it on examples that are of interest in physics.  In particular, we compute the real branch of Lambert's $W(y)$ function, which is defined as the inverse of $x \exp(x)$,  and we solve Kepler's equation. In all cases, our predictions for the theoretical errors are in excellent agreement with our numerical results, and are smaller than what could be expected from the general error analysis of spline interpolation by many orders of magnitude, namely by an astonishing $3\times 10^{-22}$ factor for the computation of $W$ in the range $W(y)\in [0,10]$, and by a factor $2\times 10^{-4}$ for Kepler's problem. In our tests, this scheme is much faster than Newton-Raphson’s method, by a factor in the range $10^{-4}$ to $10^{-3}$ for the execution time in the examples, when the values of the inverse function over an entire interval or for a large number of points are requested. For Kepler's equation and tolerance $10^{-6}$ rad, the algorithm outperforms Newton's method for all values of the number of points $N\ge 2$.
\end{abstract}



\maketitle

\section{\label{sec:introduction}Introduction}

Many problems in science and technology require the inversion of a known nonlinear function $f(x)$.  Widely studied examples include the inversion of elliptic integrals \cite{Fukushima2013,Boyd2015}, the computation of Lambert W function \cite{Corless1996,Veberic2012}, and the solution of Kepler's equation for the orbital motion of a body in a gravitational field \cite{Prussing2012,Curtis2014}. 

In many cases, the inverse function cannot be found analytically, and numerical methods must be used. Besides possible special procedures that may be found for specific forms of $f(x)$, the most popular numerical inversion schemes are those based on the Newton-Raphson method for computing the zeros of a function \cite{Mathews1999} or some of its variants \cite{Danby1983,Danby1987,Gerlach1994,Palacios2002}. These schemes are largely universal, i.e. they can be applied to a wide class of functions and converge very rapidly, especially when the value of the inverse function at one given point  is required, rather than on an entire interval. However, they require a reasonably good first guess in order to avoid problems of convergence, which may be a nontrivial issue in some cases, such as in Kepler's problem for values of the eccentricity close to one \cite{Conway1986,Charles1998,Stumpf1999}.

The rationale behind using the Newton-Raphson method is based on the fact that solving the equation $y=f(x)$ for $x$ when the value of $y$ is given is equivalent to the problem of finding the zeros of the functions $F_y(x)\equiv f(x)-y$. If a good initial guess $x_0$ of the true value $x= f^{-1}(y)$ of the zero is available, the zeros of $F_y$ can be computed by recursively applying the equation $x_{k+1}=x_{k}-\frac {F_y(x_{k})}{F_y'(x_{k})},$ i.e. ${\displaystyle x_{k+1}=x_{k}+{\frac {y-f(x_{k})}{f'(x_{k})}}}$. 

Here, a \textit{Fast Switch and Spline Inversion}  (FSSI) scheme   is described that does not require an initial guess and can be applied under very general conditions provided the function $f$ is one-to-one. The basic idea underlying this method is remarkably simple, yet it can be turned into a very powerful and accurate tool, as shall be demonstrated. Surprisingly, to our knowledge, this scheme has not been explored in the published literature. Perhaps, this may be due to an underappreciation of its rate of convergence, given the known bounds on the precision of spline interpolation, and to the existence of standard alternatives such as Newton's method.

After describing the FSSI  method, we derive theoretically a set of analytical expressions of its error estimates, and show that they are much smaller than the limit that could be derived by merely applying the existing spline analysis to this case. To optimize the algorithm, we also designed a specific spline routine that makes the FSSI more accurate and much faster than using the known spline routines. We then test the scheme on several nonlinear functions, and demonstrate that in all cases our theoretical predictions for the errors are in excellent agreement with the numerical computations.

Based upon this error analysis and on the numerical computations, the FSSI   is shown to be a valid alternative to the Newton-Raphson method (and similar quasi-Newton minimization methods) for computing values of inverse functions, especially if a good first approximation is difficult to obtain. Moreover, the FSSI   is shown to be superior to Newton-Raphson when the values $ f^{-1}(y)$ of the inverse function are required for many different $y$ points, or over an entire interval.  In the case of Kepler's equation for orbital motion, FSSI   is faster than Newton and quasi-Newton methods when the position of the orbiting body must be known at more than a few different instants, depending on the eccentricity $\text{e}$ and the requested precision. For example, for $\text{e}=0.8$ and accuracy $\sim10^{-6}$ rad, the FSSI algorithm is already  faster than Newton's even when the computation is done at $N=2$ points, and $\sim2000$ times faster for large $N$.

\section{\label{sec:FSSI}The Fast Switch and Spline Inversion (FSSI) Scheme}

In what follows the FSSI  method is described. Let $f(x)$ be the input function, which is presumed to be single valued (monotonic) in a given domain $x\in \left[x_\mathrm{min},x_\mathrm{max}\right]$.   The function $f(x)$ is assumed to be given analytically, but the case when it is known at discrete points shall also be considered.  The goal of the method is to obtain a numerical approximation for the inverse function $g(y)=f^{-1}(y)$ in the co-domain.

The FSSI  consists of a two step approach. First, when the input function is given analytically, the values of $f$ on a given grid of points $x_j$, for $j=1,\cdots,n$, are computed to obtain the matrix $(x_j,y_j)$, where $y_j=f(x_j)$. The matrix $(y_j,x_j)$, obtained by switching the arrays, gives then the \emph{exact} values $g(y_j)=x_j$ of the inverse function on the grid $y_j$. From this modified matrix, the cubic spline interpolant $S(y)$ of $(y_j,g(y_j))$ is computed by using a special routine that is designed in the next section. The resultant function $S(y)$ is the approximation of the inverse function.

FSSI  can also be used when the function $f(x)$ is specified on a grid by a set of tuples $(x_j,y_j)$. In a high-level programming language such as Python, this tuple array is represented as $(x,y)$, and the FSSI  interpolant $S(y)$ can be obtained by calling a cubic spline \emph{routine} of the switched tuple array,  $S=CubicSpline(y, x)$. In this way,  the object S in a high-level computer language would act as a generator for points in the co-domain of $f$, giving the inverse $f^{-1}$. 

\begin{figure}[!th]
\centering
\includegraphics[scale=0.75]{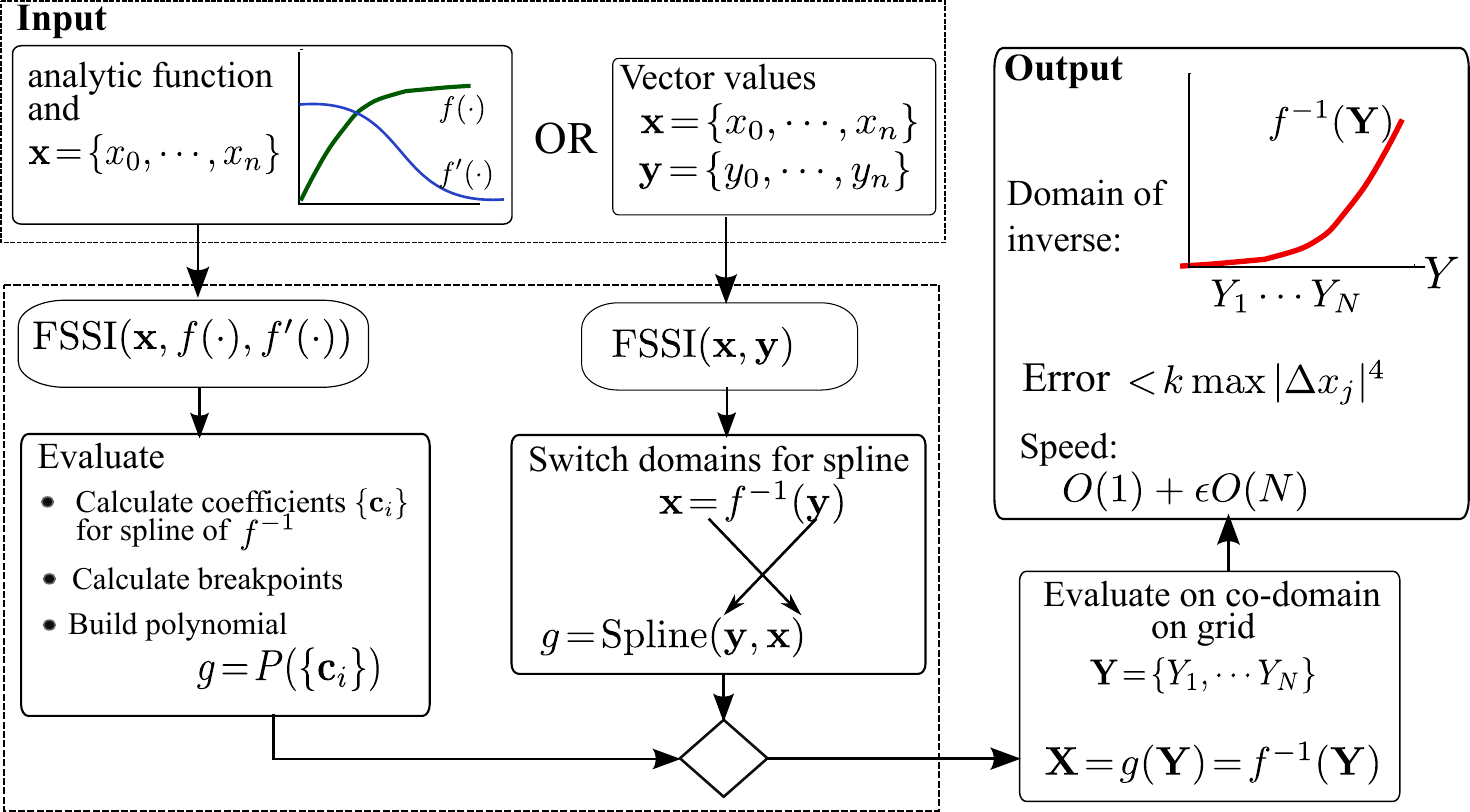}
\caption{Flow diagram of the FSSI   method for obtaining the function inverse $f^{-1}$.  The diagram indicates  the key steps of the method, as well as how it is interfaced to the input and output. }

\label{fig:figAlg}
\end{figure}

Figure \ref{fig:figAlg} shows the flow of the FSSI  algorithm, that could be implemented in any high-level computer language.  The central dotted box shows the two-step procedure of FSSI, while the outer boxes show possible  interfacing between input and output.  In particular, the input interface could accept two types of data: (1) a pointer to the analytic function $f(\cdot)$ and its derivative $f'(\cdot)$, together with a grid of $n+1$  points $x$,  or (2) the discrete tuple arrays $(x,y)$, for the case when the function or its derivative are not known explicitly.  At the output, the procedure returns a generating function, whose precision as an approximation of $f^{-1}$ is determined by the number of points $n+1$ of the input grid,  and which is used for sampling $N$ points $\{Y_1, \cdots, Y_N\}$ in the function’s   co-domain.  In subsequent sections,  we show that the FSSI   algorithm has an error bound proportional to $|k\Delta x|^4$, for k a constant (described in the text), and with a maximal time complexity of $O(1) + \epsilon\, O(N)$, where, beyond some value of $N$, the second term dominates.

\section{\label{sec:newspline}Design of a specific ultra-fast spline for the FSSI scheme}

In many problems, the function $f$ to be inverted is known analytically, along with its derivative $f'$. This is the case for Kepler's equation and for all the other examples that we will consider hereafter. Therefore, we can profit from the knowledge of $f'$ to design a specific cubic spline $S(y)$ interpolant for the FSSI algorithm.

The resulting spline makes the FSSI algorithm more  accurate and much faster than calling the spline routines that are currently available, which do not make use of the derivatives of the input function. This huge difference in speed is due to the fact that most spline routines require the numerical solution of a system of $4n$ coupled equations to compute the $4n$ coefficients of the spline \cite{Mathews1999}, where $n$ is the number of grid intervals. An exception is Akima's cubic spline \cite{Akima,ScipyAkima}, which is fast, diagonal and regular. 

The specific spline that is designed here is based on a similar idea to Akima's, but it is significantly more accurate than the latter, usually by three orders of magnitude in the examples that we shall consider, and it is also faster. Its superior performance is due to the fact that it uses the derivative $f'$ as an input, unlike Akima's. Of course, the usual applications of splines are not meant for cases in which the function to be interpolated and its derivative are given analytically. However, the situation is completely different in the problem of the inversion of a function $f(x)$. In this case, the derivatives  $g'(y_j)=1/f'(x_j)$ can be given on a grid, while the values $g(y)$ are not known. 

Let us build the specific spline $S(y)$ piecewise in each interval, $S(y)=S_j(y)$ for $y_{j}<y<y_{j+1}$, where $j$ takes the values $j=0,\cdots,n-1$. If we define the arrays $\mathbf{y_0}\equiv (y_0,\cdots,y_{n-1})$ and $\mathbf{y_1}\equiv (y_1,\cdots,y_{n})$, obtained by removing the last and the first point of the $\mathbf{y}$ array, respectively,
then the $j$-th interval can also be written as $y_{0_j}<y<y_{1_j}$. In this interval, the cubic spline can be defined as
\begin{equation}\label{Sj(y)}
    S_j(y)=\sum_{q=0}^3 c_{q_j} \,\left(y-y_{0_j}\right)^q,
\end{equation}
where for each value of $q=1,\cdots,4$  the coefficients $c_{q_j} $ can also be thought as the $n$ components of an array $\mathbf{c_q}$.

Since the values of the derivative $f'(x_j)$ of the input function are known on the grid points, we can construct an array $\mathbf{d}$ whose $n$ components are the derivatives of the inverse function $g$ on the points $y_j=f(x_j)$, 
\begin{equation}
    d_j\equiv g'(y_j)=\frac{1}{f'(x_j)},\qquad\qquad  \text{for }\qquad j=0,\cdots, n.
\end{equation}

As we did for $\mathbf{y}$, which was used to generate the arrays $\mathbf{y_1}$ and $\mathbf{y_2}$ by removing one end point, it is convenient to define similar arrays of $n$ components also from $\mathbf{x}$ and $\mathbf{d}$, namely $\mathbf{x_0}\equiv (x_0,\cdots,x_{n-1})$, $\mathbf{x_1}\equiv (x_1,\cdots,x_{n})$, $\mathbf{d_0}\equiv (d_0,\cdots,d_{n-1})$, and $\mathbf{d_1}\equiv (d_1,\cdots,d_{n})$. 
With this convention, we have to choose the spline coefficients that lead to the best approximation of the inverse function $g(y)$. The most natural choice is to force $S_j$ to coincide with the known values of the inverse function, $x_{0_j}$ and $x_{1_j}$, at the end points, and to ask the same for the derivatives $d_{0_j}$ and $d_{1_j}$. In other words, the conditions to be imposed in each interval are,

\begin{equation}\label{conditions-on-S_j}
    S_j(y_{0_j})=x_{0_j},\qquad S_j(y_{1_j})=x_{1_j},\qquad  S_j'(y_{0_j})=d_{0_j},\qquad S_j'(y_{1_j})=d_{1_j},
\end{equation}
where $S_j(y)$ is given by equation \ref{Sj(y)}.

For every fixed value of $j$, these conditions give a system of four equations for the four unknown coefficients $c_{0_j}$, $c_{1_j}$, $c_{2_j}$ and $c_{3_j}$, which is decoupled from the similar systems of equations corresponding to different values of $j$. As we have mentioned above, this is an important advantage, as compared with most of the other cubic spline interpolation methods, which must solve systems of $4 n$ coupled equations to compute the coefficients \cite{Mathews1999}, with the exception of Akima's. We can expect that this will make the FSSI algorithm using this spline much faster than using the alternative ones, and this is also what we have found numerically.

In fact, the system of equations (\ref{conditions-on-S_j}) for a fixed $j$ can be solved analytically in a straightforward way, and then implemented numerically in a completely diagonal form. In order to write the solution in a compact form, we will use a convention for vector arrays that is common in computer languages like python: an equation for arrays is interpreted in terms of components in such a way that an equality like, e.g., $\mathbf{u}=\frac{\mathbf{v}*\mathbf{w}+2*\mathbf{z}}{\mathbf{s}}$ between vectors having the same number of elements represents the equations ${u}_j=\frac{{v}_j{w}_j+2{z}_j}{s_j}$ for every $j$. With this convention, equations (\ref{conditions-on-S_j}) give the solution
\begin{align}
&\mathbf{c_0}= \mathbf{x_0},\nonumber \\
&\mathbf{c_1}= \mathbf{d_0},\nonumber \\ 
&\mathbf{c_2}=\frac {(2*\mathbf{d_0}+  \mathbf{d_1} )*( \mathbf{y_0} -  \mathbf{y_1})  - 3*( \mathbf{x_0} - \mathbf{x_1})} {( \mathbf{y_0} -  \mathbf{y_1})^2},\nonumber  \\
&\mathbf{c_3}=\frac { (\mathbf{d_0} +  \mathbf{d_1})*( \mathbf{y_0} -  \mathbf{y_1}) - 2*(\mathbf{x_0} - \mathbf{x_1})} {( \mathbf{y_0} -  \mathbf{y_1})^3}.
\label{coefficients-ourcubicspline}   
\end{align}

This result can be used when the function $f(x)$ and his derivative are known on the whole interval in which the inversion is required. If the second derivative is also known, by adding the additional conditions $ S_j''(y_{0_j})=d_{0_j}$ and $ S_j''(y_{1_j})=d_{1_j}$ to equation (\ref{conditions-on-S_j}) we can also design a diagonal quintic spline, and if also the third derivatives are known the additional conditions $ S_j'''(y_{0_j})=d_{0_j}$ and $ S_j'''(y_{1_j})=d_{1_j}$ allow for the construction of a septic spline. We have done this in both cases for the FSSI algorithm, and checked in the examples that, for a given accuracy, the resulting versions of the method perform slightly worse than with the cubic spline as designed above. Therefore, the latter will be taken as the optimal specific spline for FSSI.

\section{\label{sec:theorerr}Computation of the theoretical error}

In this section, the predicted theoretical error of the FSSI   is developed for the case of an  input function $f$ that is continuous and having continuous derivatives up to at least the fifth degree. 

Note that this error analysis not only works for the cubic spline that we have designed in the previous section, but also holds when the FSSI method is implemented with most known cubic spline routines. The main differences between the use of a cubic spline routine or the other are the speed and the accuracy very close to the end points of the $y$ domain. In both these aspects, the FSSI performs better with the spline of section \ref{sec:newspline} than with the others.

\subsection{Derivation of an upper bound on the error of the FSSI by using the known analysis of cubic spline interpolation}

Following Ref. \cite{Sonneveld1969,Mathews1999},  we can compute an upper bound for the error of the cubic spline $S(y)$,  used to interpolate the function $g(y)$,  from the formula
\begin{equation}
\vert g(y)-S(y)\vert\le \frac{1}{384}
M\, \mu,
\label{Max-error-usual}
\end{equation}
where
$M=\max\limits_{y_\mathrm{min}\le y\le y_\mathrm{max}} \left\vert g^{(4)}(y)\right\vert$, and $\mu = \max\limits_{0\le j\le n-1}(y_{j+1}-y_j)^4$.

In our case, $g(y)$ is the inverse of the input function $f(x)$, therefore it is convenient to express this error in terms of $f(x)$ and its derivatives. The $M$ term becomes,
\begin{equation}
M=
\max\limits_{x_0\le x\le x_n}
\left\vert
-\frac{15 f^{\prime\prime}(x)^3}{f^\prime (x)^7}
+\frac{10 f^{(3)}(x) f^{\prime\prime}(x)}{f^\prime(x)^6}
-\frac{f^{(4)}(x)}{f^\prime(x)^5}
\right\vert.
\label{M-in-terms-of-f}
\end{equation}

Equations (\ref{Max-error-usual}) and (\ref{M-in-terms-of-f}), along with the definition of $\mu$, can be used to obtain an upper limit on the error. As shown in examples below, the actual errors are several orders of magnitude smaller than this upper bound. In other words, the method converges much more rapidly than expected. Therefore, it is of great interest to obtain a more accurate, albeit approximate, analytical estimate of the error of FSSI  , and check its consistency in examples. This is done in the next subsection.

\subsection{Ab initio derivation of an improved estimation of the error for the FSSI}

Let us assume that the function $g(y)$ is infinitely differentiable. Thus, it can be expanded in a Taylor series $g(y)=\sum_{q=0}^{\infty}\frac{g^{(q)}(y_j)}{q!}(y-y_j)^n$ around one of the points of the grid $y_j=f(x_j)$, chosen to be the closest grid point to $y$, so that $\vert y-y_j\vert\le \vert y-y_{j+1}\vert$ and $\vert y-y_j\vert\le \vert y-y_{j-1}\vert$.

We also assume that the cubic spline interpolation is made in such a way so that it is equivalent to the first terms of this Taylor expansion, expanded around the same point and truncated beyond cubic order, $S(y)=\sum_{q=0}^{3}\frac{g^{(q)}(y_j)}{q!}(y-y_j)^q$ . Therefore, the difference is 
\begin{equation}
\vert g(y)-S(y)\vert=\left\vert\sum_{q=4}^{\infty}\frac{g^{(q)}(y_j)}{n!}(y-y_j)^n\right\vert=
\left\vert\frac{g^{(4)}(\bar y_j)}{4!}(y-y_j)^4\right\vert  
\approx
\left\vert\frac{g^{(4)}(y_j)}{4!}(y-y_j)^4\right\vert,
\label{leading-approx}
\end{equation}
where $\bar y_j$ is an unknown intermediate point between $y$ and $y_j$, and the last approximation is expected to hold for sufficiently small values of $\vert y-y_j\vert$, which is the case when a sufficiently high number of grid points is chosen.

On one hand, the exact equality in equation (\ref{leading-approx}) can be translated in the following bound,
\begin{equation}
\vert g(y)-S(y)\vert\le
\max\limits_{y_\mathrm{min}\le \bar y\le y_\mathrm{max}}\left\vert\frac{g^{(4)}(\bar y)}{4!}\right\vert
\max\limits_{0\le j\le n-1}\left\vert\frac{y_{j+1}-y_j}{2}\right\vert^4=\frac{1}{384}
M\, \mu,
\label{M-in-terms-of-f-approx}
\end{equation}
which does not use the information that $g$ is the inverse function of $f$, and coincides with the limit of equations (\ref{Max-error-usual}) and (\ref{M-in-terms-of-f}). The factor $2$ dividing the interval $y_{j+1}-y_j$ is due to the fact that $y_j$ was chosen as the closest grid point to $y$, so that $\left\vert{y-y_j}\right\vert\le \left\vert\frac{y_{j+1}-y_j}{2}\right\vert$.

On the other hand, equation (\ref{leading-approx}) can be elaborated further and expressed in terms of the function $f(x)$,
\begin{equation}\label{approx-error-full}
\vert g(y)-S(y)\vert\approx
\frac1{4!}\left\vert\left[
-\frac{15 f^{\prime\prime}(x_j)^3}{f^\prime (x_j)^7}
+\frac{10 f^{(3)}(x_j) f^{\prime\prime}(x_j)}{f^\prime(x_j)^6}
-\frac{f^{(4)}(x_j)}{f^\prime(x_j)^5}
\right]
\left[f^\prime(x_j)(x-x_j)\right]^4
\right\vert,\nonumber
\end{equation}
where $x=g(y)$ and $y-y_j\approx f^\prime (x_j)(x-x_j)$.  The last approximation is expected to be accurate over the entire interval provided the following condition
\begin{equation}
\max\limits_{0\le j\le n-1}\vert x_{j+1}-x_j\vert
\max\limits_{x_0\le x\le x_n}
\left\vert\frac{f''(x) 
}{2f'(x)}\right\vert
\ll 1
\label{approx-error-condition}
\end{equation}
is satisfied.

Assuming that the point $x_j$ is the one closest to $x$, so that $\left\vert{x-x_j}\right\vert\le \max\limits_{0\le j\le n-1}\left\vert\frac{x_{j+1}-x_j}{2}\right\vert$, the following estimation is obtained
\begin{align}\label{approx-error-split}
\vert g(y)-S(y)\vert\lessapprox & \frac{1}{384}\max\limits_{0\le j\le n-1}
\left\vert
x_{j+1}-x_j
\right\vert^4\times \\
&\max\limits_{x_0\le x\le x_n}
\left\vert
-\frac{15 f^{\prime\prime}(x)^3}{f^\prime (x)^3}
+\frac{10 f^{(3)}(x) f^{\prime\prime}(x)}{f^\prime(x)^2}
-\frac{f^{(4)}(x)}{f^\prime(x)}
\right\vert
.\nonumber
\end{align}

As described in the next section, this error approximation is usually much smaller, possibly by many orders of magnitude, than the limit of equations (\ref{Max-error-usual}) and (\ref{M-in-terms-of-f}) that we derived using existing literature on cubic spline interpolation. In fact, equation (\ref{approx-error-split}) can be expected to be a good approximation if the number of grid points $n$ is large enough and it satisfies the condition (\ref{approx-error-condition}). The  examples in the next section show that this is indeed the case.

Finally, we note that, for an equally spaced $x_j$ grid, we can substitute $\max\limits_{0\le j\le n-1}
\left\vert x_{j+1}-x_j \right\vert =\frac{x_n - x_0}{n}$ in the condition (\ref{approx-error-condition}), and
$\max\limits_{0\le j\le n-1}
\left\vert
x_{j+1}-x_j
\right\vert^4
=\frac{(x_n - x_0)^4}{n^4}$ in the estimation of the error (\ref{approx-error-split}). Not only is this the simplest choice, if the grid is not given otherwise, but it is usually also the best option.  In fact, when $\left\vert x_{j+1}-x_j \right\vert^4=\frac{(x_n - x_0)^4}{n^4}$ for every $j$,
the grid values of the inverse function, where $g$ is known exactly, are also equally spaced. As a result, the error is distributed uniformly across  the entire interval (as seen in the examples below).

\section{\label{sec:examples}Examples}

Here, the FSSI   is applied to examples of interest involving nonlinear functions $f(x)$, defined over a domain $x_0\le x\le x_n $. For the numerical computation, we developed a python code that implements the FSSI   as well as other Newton-based function inverse solvers. 

Once the numerical interpolation $S(y)$ of the inverse function is obtained, the numerical errors are computed by evaluating $S(y)-S(f(S(y)))$. In the first example, in which the exact inverse function $g(y)$ is known, we also provide an additional evaluation of the error by computing the difference $S(y)-g(y)$. In both cases, we use a grid $Y_k$ that contains 10 times as many points as the original grid $y_j = f(x_j)$. In fact, by construction $S(y)$ is exactly equal to $g(y)$--within machine errors--over the original grid $y_j$, so it is important to ensure that $S(y)$ is compared with $g(y)$ in between the grid points $y_j$. It is true that even if they were equal in number the points $Y_k$, chosen to be equally spaced, would not coincide in general with the $y_j$, whose spacing is variable and roughly proportional to $f^\prime(g(y))$; however, the election of  ten times more points  is more conservative. In this way, if the number of original grid points is large enough, a reasonable evaluation of the error is guaranteed. In fact, the examples show that when the exact analytic inverse function $g(y)$ is known, the difference $S(y)-g(y)$ has the same behaviour and magnitude of oscillations as $S(y)-S(f(S(y)))$. Moreover, the estimates are also in excellent agreement with the theoretical predictions for the error from equation (\ref{approx-error-split}).

In all the examples,  we use an equally spaced input grid in $x$, which is expected to be the best choice in most cases, as we have discussed in the previous section. The spline routine used to implement the FSSI scheme is the specific one we have designed in section \ref{sec:newspline}. However, we have also checked that similar results are obtained by calling other splines routines that do not take the derivatives of $f$ as an input, such as Scipy cubic spline routines \cite{deBoor1978, SciPyCubicSpline}. The errors in the bulk of the $y$ interval with most of those routines are very similar to each other, except very close to the boundary points, where they can be  larger by an order of magnitude than those obtained using our specific spline. An exception is Akima routine \cite{ScipyAkima}, which is less accurate by three orders of magnitude in the bulk of the interval.  This is an additional reason for preferring our specific spline, besides the fact that it is the fastest one.

\subsection{Exponential}

The first function considered is $f(x)=\exp(x)$. Of course, in this case the exact inverse function is known analytically, $g(y)=\ln(y)$, thereby serving as a validation check of our scheme.

For this case, the quantities $M$ and $\mu$,  from the bounds expressions of (\ref{Max-error-usual}) and (\ref{M-in-terms-of-f}), can be readily computed.  
The results are,
\begin{equation}
M=\max\limits_{x_0\le x\le x_n}6 e^{-4 x}=6 e^{-4 x_0},
\end{equation}
and
\begin{equation}
\mu= \left( \frac{x_n-x_0}{n}\right)^4
\max\limits_{x_0\le x\le x_n} e^{4x}
= \left( \frac{x_n-x_0}{n}\right)^4
e^{4x_n },
\end{equation}
so that the bound on the error as computed from equations (\ref{Max-error-usual}) and (\ref{M-in-terms-of-f}) is
\begin{equation}
\vert g(y)-S(y)\vert\le \frac{6}{384}\left( \frac{x_n-x_0}{n}\right)^4
e^{4(x_n-x_0) }.
\label{old-theor-error-exp}
\end{equation}

On the other hand, the analytic estimation we derived in Equation (\ref{approx-error-split}) gives
\begin{equation}
\vert g(y)-S(y)\vert\lessapprox \frac{6}{384}\left( \frac{x_n-x_0}{n}\right)^4.
\label{our-theor-error-exp}
\end{equation}

Therefore, this analytic error approximation of FSSI   is smaller by a factor $e^{-4(x_n-x_0) }$ as compared to the limit that was derived in equation  (\ref{old-theor-error-exp}) by applying the standard cubic spline error bound. For example, if $x_n-x_0=10$, then our error estimation is a factor $\exp(-40)$, i.e. 17 orders of magnitude, smaller than what could be expected from the literature. In order to benefit by this accuracy improvement, the grid must be chosen in such a way that the condition (\ref{approx-error-condition}) is satisfied, i.e.
\begin{equation}
\frac{x_n- x_0}{n}
\left\vert\frac{f''(x) 
}{2f'(x)}\right\vert
=\frac{x_n- x_0}{2 n}
\ll 1.
\label{approx-error-condition-exp}
\end{equation}

If this condition on the number of grid points $n$ is met, our improved estimation of the error  (\ref{our-theor-error-exp}) can be expected to be a good approximation. For instance, if $x\in [0,10]$, the condition becomes $n\gg 5$, so that values of $n$ of the order of 50 or larger could be sufficient. This is also what we have observed by performing numerical computations for different values of $n$. In general, for $n\gtrsim 50$, equation (\ref{our-theor-error-exp}) gives a  correct estimate for the error over the entire interval.

\begin{figure}[!th]
\centering
\includegraphics[scale=0.55]{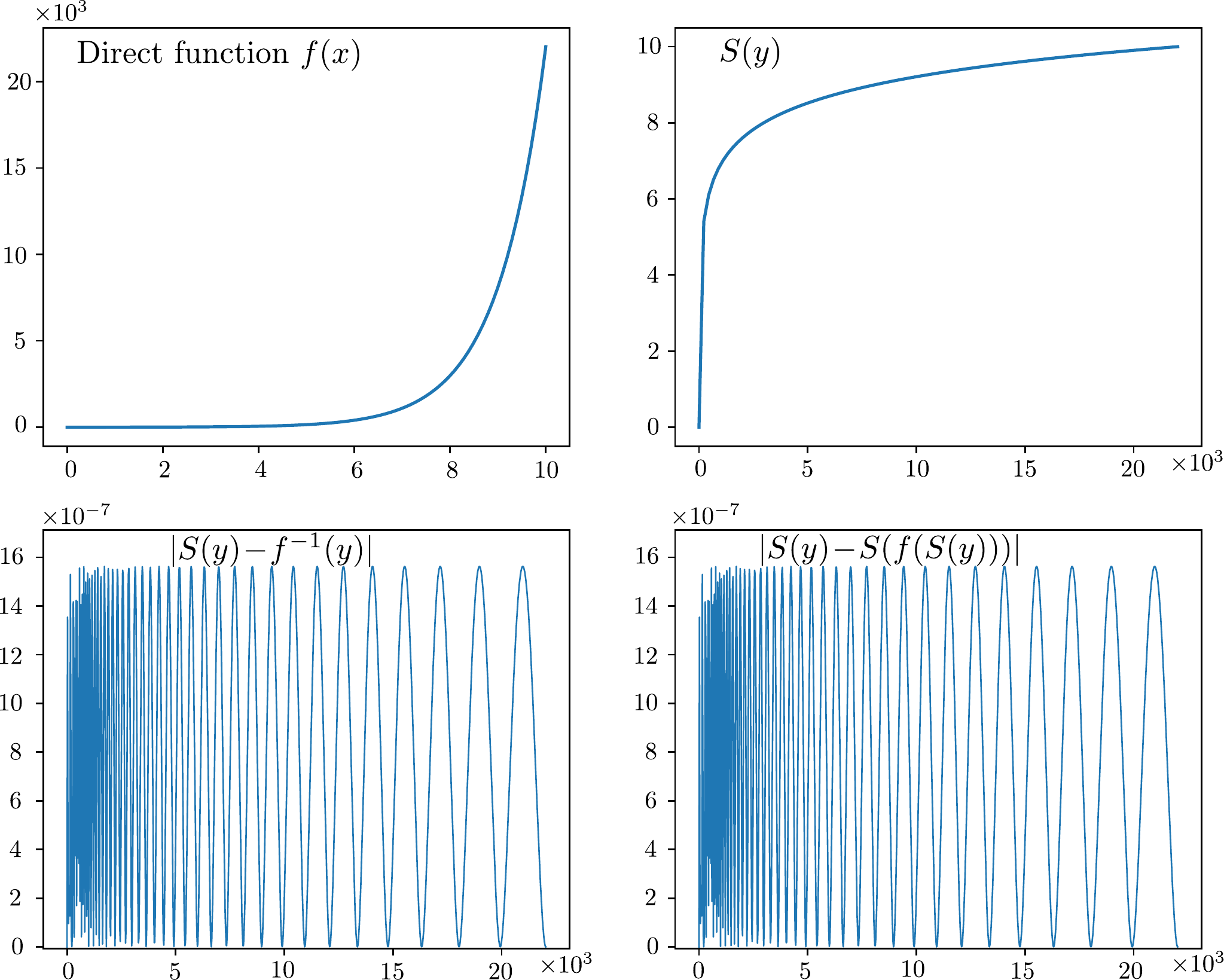}
\caption{Result of the FSSI   applied to the function $f(x)=\exp(x)$ (top left) over the domain $x\in [0,10]$. The FSSI   interpolant $S(y)$ is shown for $n=10^2$ grid points (top right), together with two independent evaluations of the  numerical errors: i) $\vert S(y)-f^{-1}(y)\vert$, where $f^{-1}(y)=\ln(y)$ (bottom left); ii)  $\vert S(y)-S(f(S(y)))\vert$ (bottom right).}
\label{fig:fig1}
\end{figure}

Figure \ref{fig:fig1} shows the result of the FSSI   for the inversion of $f(x)=\exp(x)$ over the domain $x\in [0,10]$ using $n=10^2$ grid points. In this case, our theoretical prediction of Equation (\ref{our-theor-error-exp}) gives $\vert g(y)-S(y)\vert\lessapprox 1.6\times 10^{-6}$, which is in excellent agreement with the numerical computation over the entire interval. The results for this case  also confirm the theoretical prediction that  FSSI   is 17 orders of magnitude more accurate than what could be expected by naively applying the general results for cubic splines, as in equation (\ref{old-theor-error-exp}). An important feature of Figure \ref{fig:fig1} is that the error is distributed uniformly across the interval. As discussed previously, this is a consequence of choosing an equally spaced grid for $x$.

\subsection{Lambert W function}

Let $f(x)=x\exp(x)$, whose inverse function $g(y)$ in the real domain is the principal branch of Lambert's W function, $W(y)$ \cite{Corless1996,Veberic2012}. In this case, the FSSI   interpolation $S(y)$ can be compared with the values of $W(y)$ that are computed with other methods. The values of $M$ and $\mu$ from equations (\ref{Max-error-usual}) and (\ref{M-in-terms-of-f}) for the theoretical bound are,
\begin{equation}
M=\max\limits_{x_0\le x\le x_n}\left| -\frac{15 \left(e^x x+2 e^x\right)^3}{\left(e^x x+e^x\right)^7}+\frac{10 \left(e^x x+3 e^x\right) \left(e^x x+2 e^x\right)}{\left(e^x x+e^x\right)^6}-\frac{e^x x+4 e^x}{\left(e^x x+e^x\right)^5}\right|,
\end{equation}
which is a monotonically decreasing function for $x>-1$, and
\begin{equation}
\mu= \left( \frac{x_n-x_0}{n}\right)^4
\max\limits_{x_0\le x\le x_n} \left| e^x x+e^x\right| ^4,
\end{equation}
which increases monotonically. Therefore, the bound (\ref{Max-error-usual}) becomes
\begin{equation}
\vert g(y)-S(y)\vert\le \frac{1}{384}\,\mu(x=x_n)\,M(x=x_0).
\label{old-theor-error-W}
\end{equation}

For example, over the domain  $x\in [0,10]$ this gives
$\vert g(y)-S(y)\vert\lessapprox 5.7\times 10^{20}\left( \frac{x_n-x_0}{n}\right)^4$.

On the other hand, our analytical estimation from Equation (\ref{approx-error-split}) becomes
\begin{align}\label{our-theor-error-W}
&\vert g(y)-S(y)\vert\lessapprox \frac{1}{384}\left( \frac{x_n-x_0}{n}\right)^4\times\\
&\max\limits_{x_0\le x\le x_n}
\left| -\frac{15 \left(e^x x+2 e^x\right)^3}{\left(e^x x+e^x\right)^3}+\frac{10 \left(e^x x+3 e^x\right) \left(e^x x+2 e^x\right)}{\left(e^x x+e^x\right)^2}-\frac{e^x x+4 e^x}{e^x x+e^x}\right|.
\nonumber
\end{align}

The function to be maximized in equation (\ref{our-theor-error-W})  monotonically decreases for $x>-1$, so that its maximum is achieved for $x=x_0$.   Thus,  over the domain  $x\in [0,10]$ our estimate of the error (\ref{our-theor-error-W}) gives
$\vert g(y)-S(y)\vert\lessapprox 0.17\left( \frac{x_n-x_0}{n}\right)^4$,
which is $3\times 10^{-22}$ times smaller than the bound (\ref{old-theor-error-W}) that is obtained by applying the standard spline error analysis,  as in equation (\ref{Max-error-usual}). In this case,  the condition (\ref{approx-error-condition}) for the applicability of our approximation (\ref{our-theor-error-W}) becomes
\begin{equation}
    n\gg \frac{{x_0- x_n}}{2} \max\limits_{x_0\le x\le x_n}\left\vert \frac{x+2}{x+1}\right\vert = 10,
\end{equation}
which is a surprisingly low value, for such a huge variation of $f$.

\begin{figure}[!th]
\centering
\includegraphics[scale=0.55]{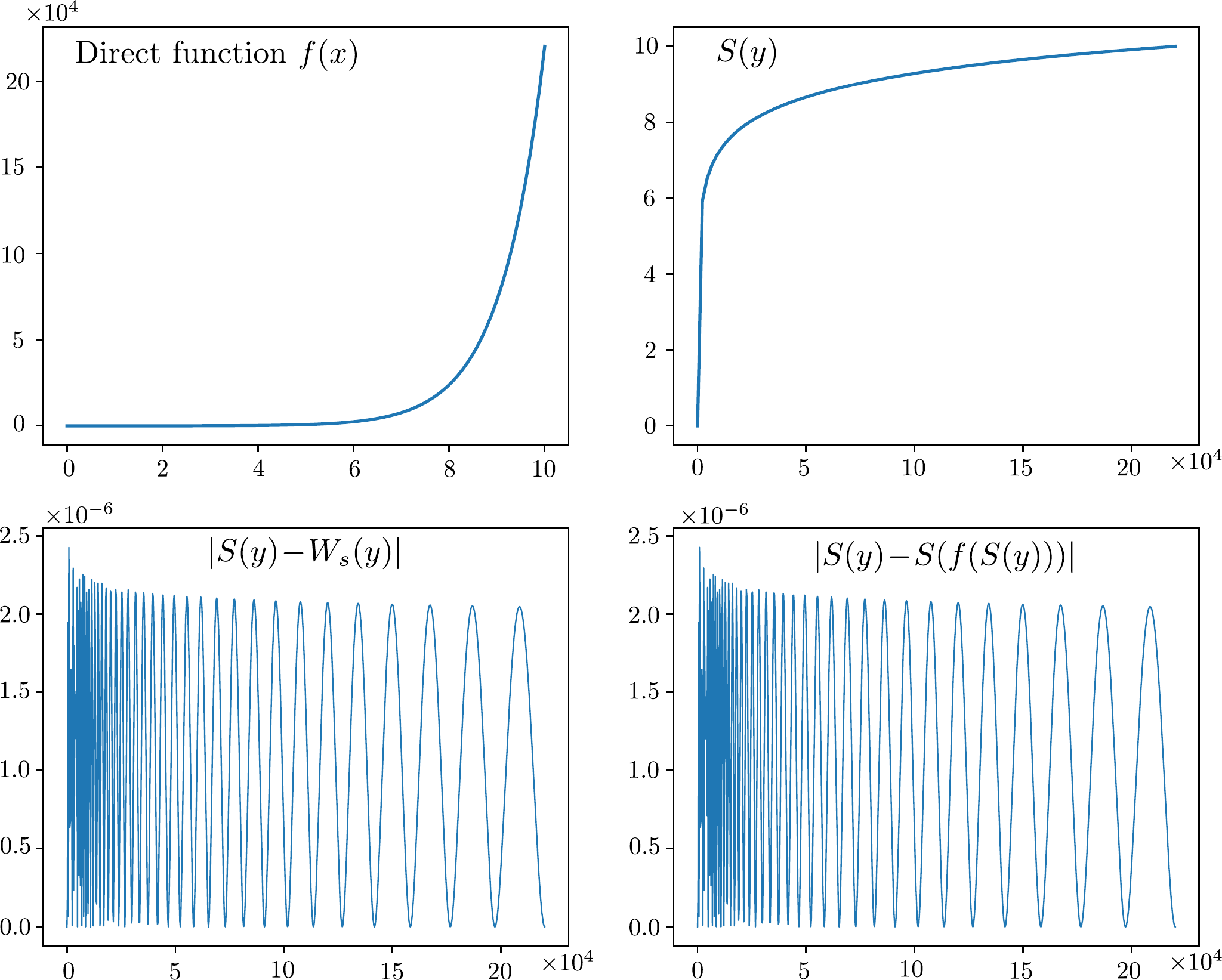}
\caption{Numerical result of the FSSI   applied to the function $f(x)=x\exp(x)$ (top left) over the domain $x\in [0,10]$. The FSSI   interpolant $S(y)$ is shown for $n=10^2$ grid points (top right), together with  two independent evaluations of the numerical errors: i) $\vert S(y)-f^{-1}(y)\vert$, where $f^{-1}(y)=W_S(y)$ as computed with scipy.special.lambertw routine (bottom left); ii)  $\vert S(y)-S(f(S(y)))\vert$ (bottom right).}
\label{fig:fig3}
\end{figure}

Figure \ref{fig:fig3} shows the numerical result of the FSSI   inversion of $f(x)=x\exp(x)$ in the domain $x\in [0,10]$ using $n=10^2$ grid points. Two independent evaluations of the numerical errors are given: i) $\vert S(y)-f^{-1}(y)\vert$, where $f^{-1}(y)=W_S(y)$ as computed with scipy.special.lambertw routine; ii)  $\vert S(y)-S(f(S(y)))\vert$. The fact that they agree with each other provides confirmation concerning our treatments of the errors. Moreover, in this case our theoretical prediction of Equation (\ref{our-theor-error-W}) gives $\vert g(y)-S(y)\vert\lessapprox1.7\times 10^{-5}$, and the numerical error not only agrees with it, but it is even much smaller, by almost an order of magnitude, $\vert g(y)-S(y)\vert_\text{numerical}<2.5\times 10^{-6}$  over the entire interval. In this case, FSSI   is more accurate by an astonishing factor $4\times 10^{-23}$ than what could be expected by naively applying the general results for cubic splines, as in equation (\ref{old-theor-error-W}).

\subsection{Kepler's equation}

Kepler's equation for an elliptical orbital motion of eccentricity $\text{e}$ can be written as
\begin{equation}
    y=x-\text{e} \sin x,
\end{equation}
where $y$ and $x$ represent the so-called mean and eccentric anomaly, respectively  \cite{Prussing2012,Curtis2014}. The former is the time elapsed since periapsis, as  measured in radians, $y=\frac{2\pi t}{T} $, where $T$ is the period of the orbit. The eccentric anomaly $x$ is related to the angle $\theta$ between the position vectors at periapsis and at time $t$, with origin in the center of gravity, through the equation
\begin{equation}
    \theta=2\arctan\left(\sqrt{\frac{1+\text{e} }{1-\text{e} }}\tan\frac{x}{2}\right).
\end{equation}

A fundamental problem in orbital dynamics  \cite{Prussing2012,Curtis2014}  is to obtain the time dependence of the angle $\theta$ describing the position of the orbiting body at time $t$, which requires the inversion of the function  $y=f(x)\equiv x-\text{e} \sin x$. Taking into account that the orbit is periodic, and that for $x\in[\pi,2\pi]$ we have $f(x)=2\pi-f(2\pi-x)$, it is sufficient to consider only the interval $x\in[0,\pi]$ to obtain the behavior for all values of $x$. The corresponding co-domain is then $y\in[0,\pi]$ \cite{Prussing2012,Curtis2014}.

The inverse function $x=g(y)$ will yield the eccentric anomaly as a function of the mean anomaly, and thus the evolution $\theta(t)$ will be obtained. This is usually done in an efficient way using Newton's method with the first guess $x_0=y+\text{e}/2$ \cite{Prussing2012,Curtis2014,Danby1983}.

Here,  FSSI   is considered as an alternative to Newton-based methods for solving Kepler's equation. In this case, the values of $M$ and $\mu$ in the theoretical bound of equations (\ref{Max-error-usual}) and (\ref{M-in-terms-of-f}) are
\begin{equation}
M=\max\limits_{0\le x\le \pi}\left| -\frac{15 \text{e}^3 \sin ^3x}{(1-\text{e} \cos x)^7}+\frac{10 \text{e}^2 \sin x \cos x}{(1-\text{e} \cos x)^6}+\frac{\text{e} \sin x}{(1-\text{e} \cos x)^5}\right|
\end{equation}
and
\begin{equation}
\mu=
\left( \frac{\pi}{n}\right)^4
\max\limits_{0\le x\le \pi}
\left| 1-\text{e} \cos x\right| ^4.
\end{equation}

As a concrete example, the case of $\text{e}=0.8$ is considered. Thus, the maximum values are $M=10275.1$, which is obtained for $x=0.166$, and $\mu=\left| 1+\text{ec}\right| ^4\left( \frac{\pi}{n}\right)^4=10.4976\left( \frac{\pi}{n}\right)^4$, obtained for $x=\pi$. Therefore  the bound (\ref{Max-error-usual}) becomes
\begin{equation}
\vert g(y)-S(y)\vert\lesssim
\frac{2.7\times10^4}{n^4}.
\label{old-theor-error-Kepler}
\end{equation}

However, the expression from our analytic estimation from Equation (\ref{approx-error-split}) becomes
\begin{equation}
\vert g(y)-S(y)\vert\lessapprox \frac{1}{384}\left( \frac{\pi}{n}\right)^4
\max\limits_{0\le x\le \pi}
\left| -\frac{15 \text{e}^3 \sin ^3x}{(1-\text{e} \cos x)^3}+\frac{10 \text{e}^2 \sin x \cos x}{(1-\text{e} \cos x)^2}+\frac{\text{e} \sin x}{1-\text{e} \cos x}\right|.
\end{equation}

For $\text{e}=0.8$, the expression in the $\vert\;\;\vert$ bracket has a maximum  value 21.586 obtained for x=0.214657, therefore we obtain
\begin{equation}
\vert g(y)-S(y)\vert\lessapprox \frac{5.5}{n^4}.
\label{our-theor-error-Kepler}
\end{equation}

Thus, our estimation for the theoretical error (\ref{our-theor-error-Kepler}) in this case is $2\times 10^{-4}$ smaller than what could be expected by naively applying the known bounds on cubic spline interpolation. Here, the condition (\ref{approx-error-condition}) for the applicability of our approximation (\ref{our-theor-error-Kepler}) becomes
\begin{equation}
  n\gg \frac{\pi\, \mathrm{e}}{2} \max\limits_{0\le x\le \pi}\left\vert \frac{\sin x}{1-\mathrm{e}\cos x}\right\vert \simeq 2.
\end{equation}
As a result, for Kepler problem, the FSSI method and the estimation (\ref{our-theor-error-Kepler}) start to be reliable for $n$ as small as the order of ten.

\begin{figure}[!th]
\centering
\includegraphics[scale=0.55]{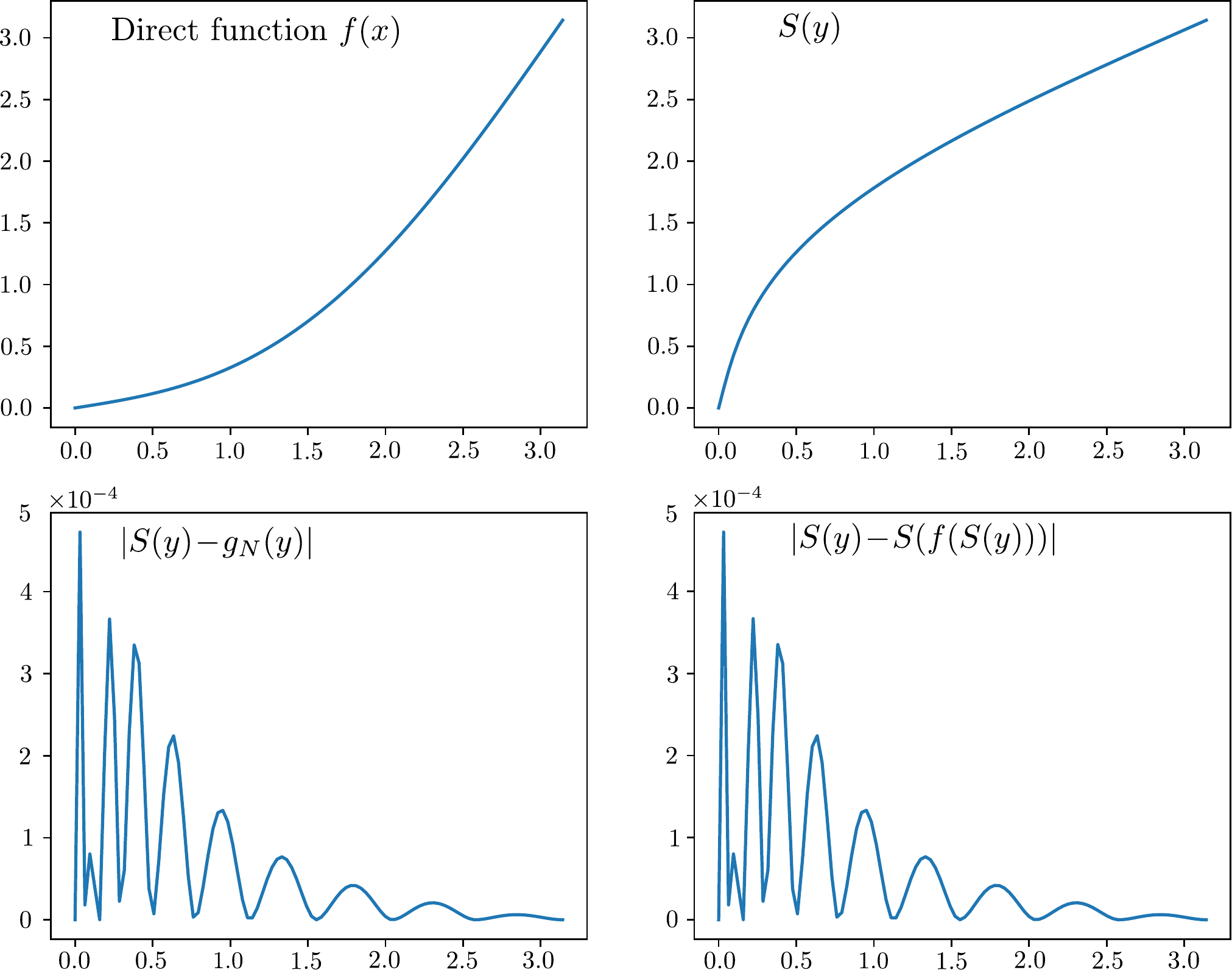}
\caption{Numerical result of the FSSI   applied to the function $f(x)=x-0.8 \sin x$ over the domain $x\in [0,\pi]$ (top left), corresponding to Kepler's equation for an elliptical orbit of eccentricity 0.8. The FSSI   interpolant $S(y)$ is shown for $n=10$ grid points (top right), together with two independent evaluations of the  numerical errors: i) $\vert S(y)-g_{\rm N}(y)\vert$, where $g_{\rm N}(y)$ is computed with Newton's method (bottom left); ii)  $\vert S(y)-S(f(S(y)))\vert$ (bottom right).}
\label{fig:fig4}
\end{figure}

\begin{figure}[!th]
\centering
\includegraphics[scale=0.55]{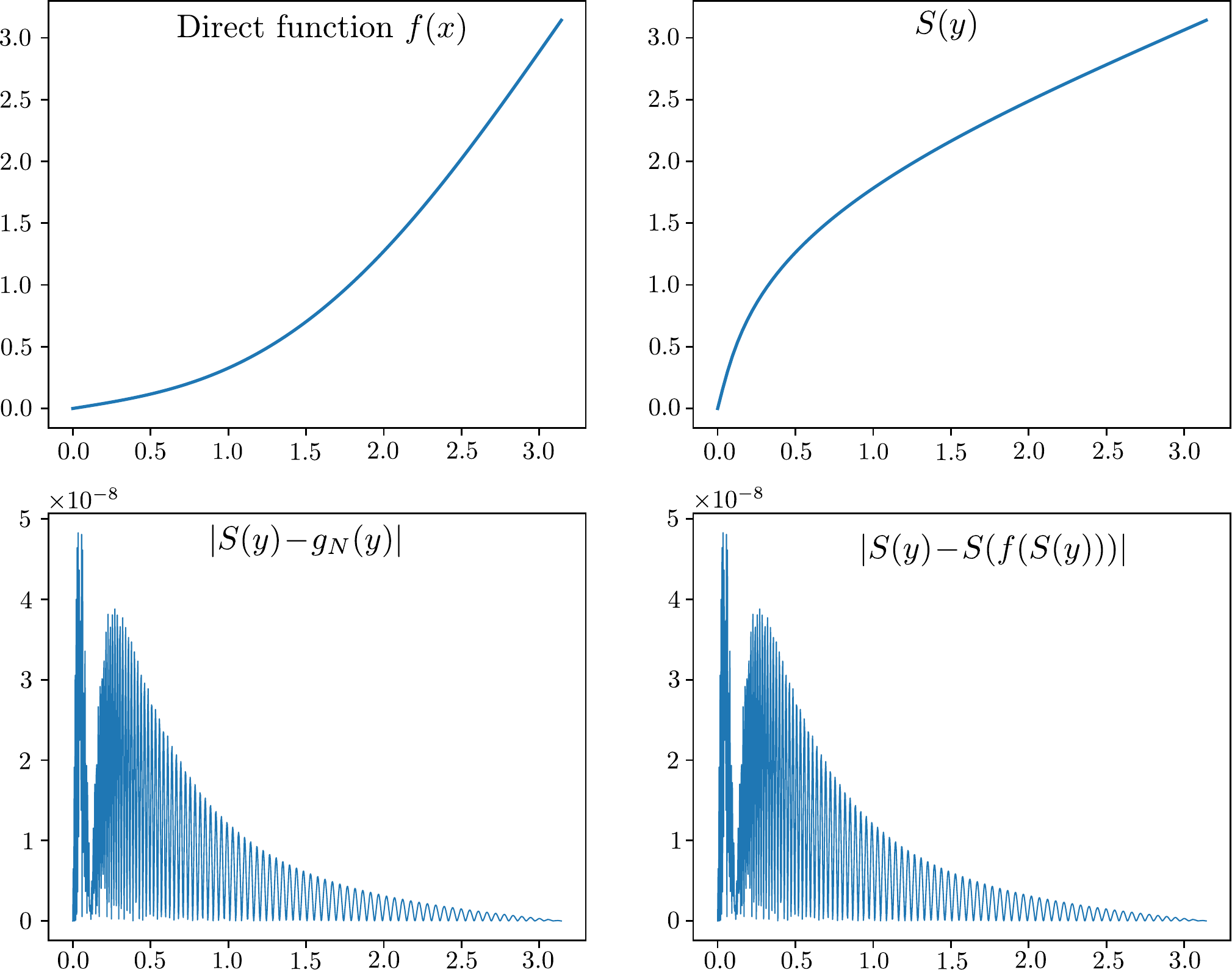}
\caption{Numerical result of the FSSI   applied to the function $f(x)=x-0.8 \sin x$ over the domain $x\in [0,\pi]$ (top left), corresponding to Kepler's equation for an elliptical orbit of eccentricity 0.8. The FSSI   interpolant $S(y)$ is shown for $n=10^2$ grid points (top right), together with two independent evaluations of the  numerical errors: i) $\vert S(y)-g_{\rm N}(y)\vert$, where $g_{\rm N}(y)$ is computed with Newton's method (bottom left); ii)  $\vert S(y)-S(f(S(y)))\vert$ (bottom right).}
\label{fig:fig4bis}
\end{figure}

Figures \ref{fig:fig4} and \ref{fig:fig4bis} show the result of the FSSI   for the inversion of $f(x)=x-0.8\sin{}x\;{}$ over the domain $x\in [0,\pi]$ using $n=10$ and $n=10^2$ grid points, respectively. In these cases, our theoretical prediction of Equation (\ref{our-theor-error-Kepler}) gives $\vert g(y)-S(y)\vert\lessapprox 5.5\times (10^{-4}\;\text{or}\;10^{-8})$, respectively, in excellent agreement with our numerical computation over the entire interval. 

Again, we provide two independent numerical computations of the error, one obtained by plotting the difference of the FSSI   interpolation with the values of $g_{\rm N}(y)$ obtained with Newton's method, and the other given by the difference $S(y)-S(f(S(y)))$. The fact that these evaluations of the error also agree with each other is a further confirmation of the validity of our error analysis. 

By comparing figures \ref{fig:fig4} and \ref{fig:fig4bis}, we also see that the accuracy scales with $n^{-4}$, as equation (\ref{our-theor-error-Kepler}) predicts, and that our estimation for the error is reliable even for just $n=10$ grid points.

\section{\label{sec:comparison}Numerical Comparisons with Newton-based methods}

Apart from the numerical calculations for error analysis,  we carried out numerical comparisons between FSSI   and  Newton-based methods (as well as the scipy.lambertw, for the case of Lambert $W$ calculation) for calculating the inverse of single-valued functions. As in the examples of the previous section, the FSSI   and Newton-based methods were implemented in the Python programming language, respecting standard practice of minimizing loops and relying upon library function calls (that depend upon compiled code). When possible, we also tested accelerating all methods with Numba JIT compilation, however we found that no considerable difference in empirical execution times could be appreciated.

The algorithms \ref{alg:alg1} and \ref{alg:alg2}  provide the steps of  the FSSI   method and the  generalized Newton-Raphson method, respectively, used in the benchmark comparisons. This simple version of Newton-Raphson method has been shown to be almost as fast as more elaborate versions, the difference in the execution times being usually below $\sim 30\%$ \cite{Palacios2002}.

\begin{minipage}{.48\linewidth}
  \begin{algorithm}[H]
  \caption{Benchmark for FSSI\label{alg:alg1}}
  \begin{algorithmic}[1]
    \Procedure{bench\_FSSI}{$\mathbf{Y},\mathbf{x}, f(x), f'(x)$}
    \State $\mathbf{y} = f(\mathbf{x})$ 
    \State $\mathbf{d} = \frac{1}{f'(\mathbf{x})}$ 
    \State $\mathbf{c_0} = \mathbf{x}[:-1]$ 
    \State $\mathbf{c_1} = \mathbf{d}[:-1]$ 
    \State $\mathbf{d1} = \mathbf{d}[1:]$ 
    \State $\mathbf{xx} = \mathbf{c_0}- \mathbf{x}[1:]$ 
    \State $\mathbf{y0} = \mathbf{y}[:-1]$ 
    \State $\mathbf{y1} = \mathbf{y}[1:]$ 
     \State $\mathbf{yy} = \mathbf{y0}- \mathbf{y1}$ 
    \State $\mathbf{yy2} =\mathbf{yy}*\mathbf{yy}$ 
    \State $\mathbf{yd1} =\mathbf{yy}* \mathbf{d1}$
    \State $\mathbf{yd0} =\mathbf{yy}* \mathbf{c1}$ 
    \State $\mathbf{c_2}=\frac {2* \mathbf{yd0}+  \mathbf{yd1}   - 3*\mathbf{xx}} { \mathbf{yy2}}$ 
    \State $\mathbf{c_3}= \frac { \mathbf{yd0} +  \mathbf{yd1} - 2*\mathbf{xx}} {\mathbf{yy2}*\mathbf{yy}}$
        \State  call \text{P}: $\mathbf{X}=\text{P}((\mathbf{c_3},\mathbf{c_2},\mathbf{c_1},\mathbf{c_0}),\mathbf{Y})$ 
        \State \textbf{return} $\mathbf{X}$ \Comment{ $\!=\! f^{-1}(\mathbf{Y})$}
   \EndProcedure
  \end{algorithmic}
\end{algorithm}
\end{minipage} \hspace{0.7cm}
\begin{minipage}{.46\textwidth} %
\begin{algorithm}[H]
 \caption{Benchmark for Newton \label{alg:alg2}}
  \begin{algorithmic}[1]
    \Procedure{bench\_Newton}{$\mathbf{Y}, f(x), f^\prime(x), \text{tol}$}
    \For{$Y_k$ in $\mathbf{Y}$}
       \State $X_k = g_0(Y_k)$
       \State $\Delta = \left|\frac{Y_k-f(X_k)}{f^\prime(X_k)}\right|$
       \While{$\Delta > \text{tol}$}
           \State $\gamma = \frac{Y_k-f(X_k)}{f^\prime(X_k) }$
           \State $X_k =X_k + \gamma$
           \State $\Delta = \left|\gamma\right|$
       \EndWhile
    \EndFor
   \State \textbf{return} $\mathbf{X}$\Comment{$\!=\! f^{-1}(\mathbf{Y})$}
   \EndProcedure
  \end{algorithmic}
\end{algorithm}
\end{minipage} 
\vspace{0.5cm}

In algorithm \ref{alg:alg2}, we have called $g_0(Y_k)$ the initial guess for Newton's method as a function of $Y$, which is to be chosen depending on the problem considered.

In algorithm  \ref{alg:alg1}, we have followed the conventions of section \ref{sec:newspline} for the arrays, which are indicated in boldface. Accordingly, the operations involving them are to be understood to be valid for the components, i.e. they run over $\{i=0,n\}$ or over  $\{i=1,n\}$, for lowercase arrays, or over $\{i=1,N\}$, for uppercase arrays. An exception are the expressions $\mathbf{v}[1:]$ and $\mathbf{v}[:-1]$, which mean the removal of the first or the last element from $\mathbf{v}$, respectively.

In Python, the piecewise polynomial function $P$, corresponding to equation (\ref{Sj(y)}), can be obtained  in terms of the breakpoints and the coefficients $\mathbf{c_q}$ using the subroutine PPoly \cite{PPoly}, so that P = PPoly. Another possibility is to write an explicit subroutine for computing the polynomial, in which the insertion points $j$ are located by binary search using scipy.searchsorted \cite{searchsorted}. The two possibilities are shown below:

\vspace{0.5cm}

\noindent
\begin{minipage}{.52\linewidth}
  Subroutine P for $\text{FSSI}\_\textsc{PPoly}$ in Python
\begin{algorithmic}[1]
    \Function{P}{$(\mathbf{c_3},\mathbf{c_2},\mathbf{c_1},\mathbf{c_0}),Y$}
        \State     P = scipy.PPoly 
     \EndFunction
\end{algorithmic}
\end{minipage}
\begin{minipage}{.52\linewidth}
Subroutine P for $\text{FSSI}\_\textsc{Search}$ in Python
\begin{algorithmic}[1]
    \Function{P}{$(\mathbf{c_3},\mathbf{c_2},\mathbf{c_1},\mathbf{c_0}),\mathbf{Y}$}
        \State     $j=\text{numpy.searchsorted}(\mathbf{y1},\mathbf{Y})$
        \State     $\mathbf{P1}=\mathbf{Y}-\mathbf{y_{0_j}}$
        \State     $\mathbf{P2}=\mathbf{P1}*\mathbf{P1}$
        \State     $\mathbf{S}=\mathbf{c_{o_j}}+\mathbf{c_{1_j}}*\mathbf{P1}+\mathbf{c_{2_j}}*\mathbf{P2}+\mathbf{c_{3_j}}*\mathbf{P2}*\mathbf{P1}$
         \State \textbf{return} $\mathbf{S}$   
      \EndFunction
\end{algorithmic}
\end{minipage}
\vspace{0.5cm}

A discrete analysis of the algorithms \ref{alg:alg1} and \ref{alg:alg2} shows that the FSSI   executes in constant time $\mathcal{O}(1)$, because once the spline coefficients are obtained with a grid given by $n$ points, all subsequent $N$ function evaluations are equivalent array access through the generating function.  However, when $N$ is large, finite cache sizes and the search of the breakpoints overtake this  behavior, so that the algorithm follows a linear time dependence $\mathcal{O}(N)$ \cite{Oded2008, Brent2010}.

In other words, the execution time can be written as
$\Delta t_\textsc{FSSI\_PPoly}\simeq\epsilon N+ \eta$ and $\Delta t_\textsc{FSSI\_Search}\simeq\beta N + \alpha$, for the python implementations of FSSI with PPoly or Searchsorted, respectively. As we show below, $\Delta t_\textsc{FSSI\_PPoly}<\Delta t_\textsc{FSSI\_Search}$ for large $N$, typically $N\gtrsim 10^4$, and  $\Delta t_\textsc{FSSI\_PPoly}>\Delta t_\textsc{FSSI\_Search}$ for lower values of $N$. 
We can then merge the two python routines for P, algorithms FSSI\_Search  and FSSI\_PPoly, by choosing the fastest one with an if statement, e.g. if $N > 10^4$ do PPoly, else do the routine with searchsorted. The execution time for this combined routine is $\Delta t \simeq \alpha+\epsilon N$, i.e. it behaves as $\mathcal{O}(1) + \epsilon\, \mathcal{O}(N)$.

On the other hand, for the Newton minimization based methods, all evaluations of the function inverse occur with an average number of iterations, $m$ (as seen in the \textit{while} loop of lines 6-10), therefore, these algorithms have a lower bound linear time behavior $\mathcal{O}(m N)$ for all values of $N$.

To obtain an empirical execution time comparison between methods, we ran the benchmarks for two cases: the calculation of the Lambert W function, and the solution of Kepler's problem.   

The details of the numerical comparison are as follows:
\begin{itemize}
\item \textit{Hardware:} The numerical comparisons were carried out on a modest desktop computer (a 64 bit Intel i5-2400 CPU \@ 3.10GHz, with 32GB memory, and with the Ubuntu/Linux operating system with 4.13.16 kernel).
\item \textit{Tolerance:} For each case the same level as the error of the FSSI   in this case:  For the Lambert W problem, we used a tolerance $2\times10^{3}/n^4$ for scipy.lambertw and Newton; For the Kepler solution, we used a tolerance $6/n^4$ for Newton and Pynverse \cite{Pynverse} quasi-Newton method.
\item
For Lambert W,  we chose the simplest first guess, $g_0(Y_k)=\frac{x_0+x_n}{2}=5$. Of course, better choices may be found, but we want to use this case to compare FSSI and Newton in the absence of a good first guess. On the other hand, we also penalize the FSSI method by taking the tolerance for Newton-based methods equal to the theoretical error of FSSI, which overestimates the numerical error by an order of magnitude as shown in section \ref{sec:examples}. 
\item
In the case of Kepler's  equation, we take $\text{e}=0.8$ and we use a very good first guess, $g_0(Y_k)=Y_k+\frac{\text{e}}{2}$, as was mentioned in section \ref{sec:examples}.
\end{itemize}

\begin{figure}[!th]
\centering
\includegraphics[scale=0.4]{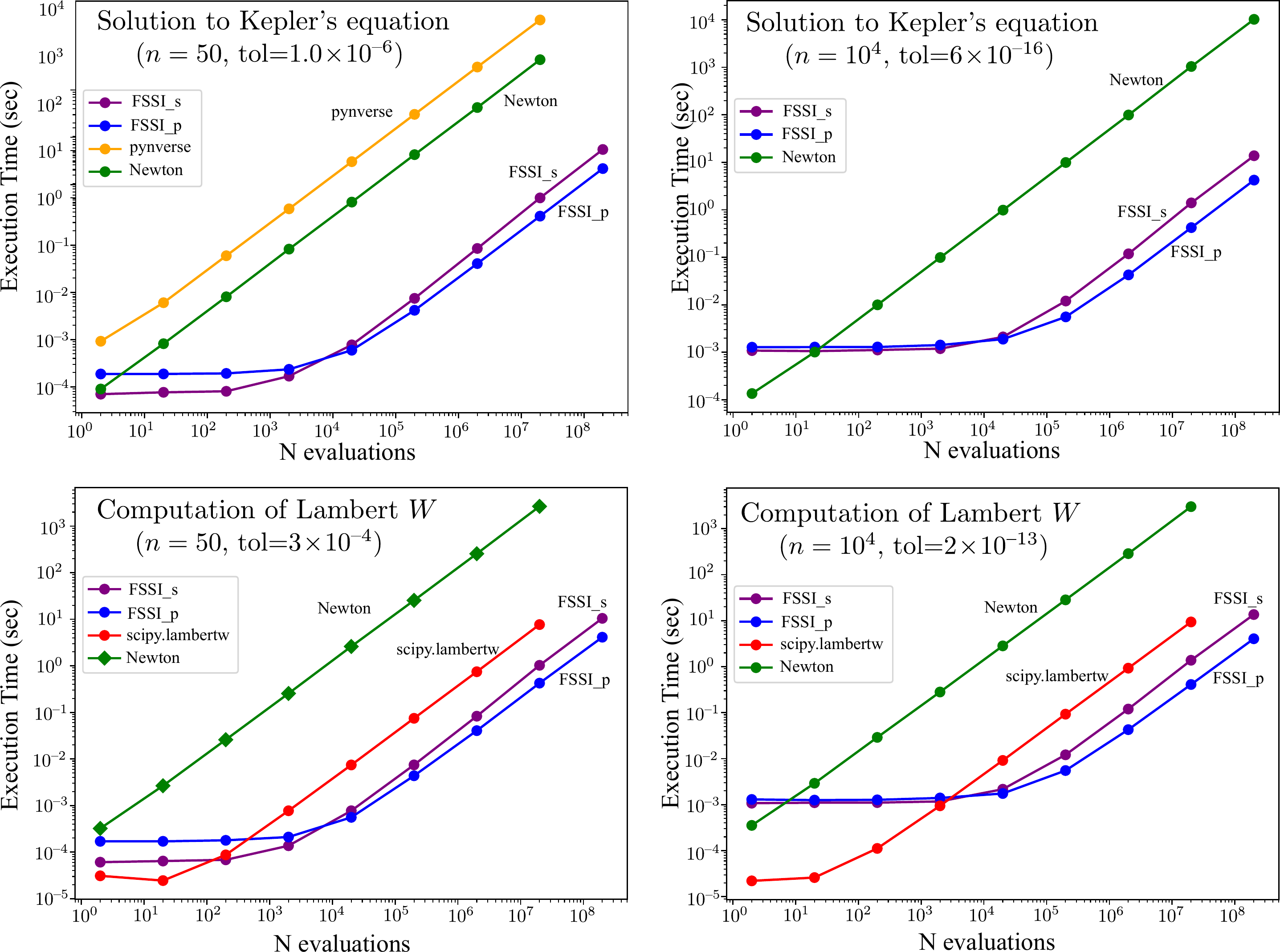}
\caption{Numerical comparisons of FSSI   and other methods for the solution of Kepler's equation (top) and computation of the Lambert W function (bottom). $\text{FSSI}\_\text{p}$ and  $\text{FSSI}\_\text{s}$  stand for the algorithm using PPoly or Searchsorted subroutines, respectively.}
\label{fig:fig5}
\end{figure}

Figure \ref{fig:fig5} shows empirical execution time comparisons between different  numerical algorithms and FSSI   for calculating Lambert W and for solving Kepler's equation.  The results support the theoretical expectations described above. For the FSSI   method,  there is a wide range of  $N$ values for which the $\mathcal{O}(N)$ behavior is negligible as compared with the $\mathcal{O}(1)$ behavior; however for very large $N$, when the $\mathcal{O}(N)$ part dominates, the linear coefficient $\epsilon$ is several orders of magnitude smaller than those of the other methods available.  

In all the cases, Pynverse \cite{Pynverse} (based on a quasi-newton optimization) is much slower than the other methods considered, which is not a surprise since it is meant to be universal, rather than fast. Therefore,  we will limit our discussion to the comparison between FSSI and Newton-Raphson methods.  As we see from  figure \ref{fig:fig5}, FSSI is not only universal, but it is also fast, and for large $N$ it is the fastest method.

These results have been used to obtain linear fits to the data.
For example, for Kepler's problem with $n=50$, corresponding to tolerance $10^{-6}$ rad (which can be a sufficient accuracy for orbit determination in many cases), we found
$\Delta t_\textsc{FSSI\_PPoly}\simeq 2.1\times 10^{-8} N + 1.9\times 10^{-4}$ and $\Delta t_\textsc{FSSI\_Search}\simeq 5.3\times 10^{-8} N   + 7.2\times 10^{-5}$. 
These values of the coefficients have been obtained by separate fits to the low $N$ data, for $\eta$ and $\alpha$, and to the high $N$ data, for $\epsilon$ and $\beta$, in order to get the best estimates in these regimes, so that the approximation is slightly worse for $10^3\lesssim N\lesssim 10^4$.

In any case, as shown in figure \ref{fig:fig5}, $\textsc{FSSI\_PPoly}$ is faster than $\textsc{FSSI\_Search}$ 
for $N\gtrsim 10^4$ and slower for $N\lesssim 10^4$.
By choosing the best of the two variants with an if statement, 
we obtain a combined behavior $\Delta t_\textsc{FSSI}\simeq 2.1\times 10^{-8} N + 7.2\times 10^{-5}$. This should be compared with the execution time for Newton-Raphson method, $\Delta t_\textsc{Newton}= 4.2\times 10^{-5} N$. We find that $\Delta t_\textsc{Newton}>\Delta t_\textsc{FSSI}$ for every $N\ge 2$, and that FSSI is $\sim2\times 10^{3}$ faster than Newton-Raphson for large $N$.

Similarly, for Kepler's problem with $n=10^4$, corresponding to tolerance $6\times 10^{-16}$ rad,  we obtain a combined behavior 
$\Delta t_\textsc{FSSI}\simeq 2.1\times 10^{-8} N + 1.1\times 10^{-3}$ while 
$\Delta t_\textsc{Newton}= 5.0\times 10^{-5} N$. We find that $\Delta t_\textsc{Newton}>\Delta t_\textsc{FSSI}$ for every $N\gtrsim 20$, and that FSSI is still $\sim 2\times 10^{3}$ faster than Newton-Raphson for large $N$.

For Lambert W with $n=50$, corresponding to tolerance $3\times 10^{-4}$, we obtain a combined behavior 
$\Delta t_\textsc{FSSI}\simeq 2.1\times 10^{-8} N + 6.2\times 10^{-5}$ while 
$\Delta t_\textsc{Newton}=1.3\times 10^{-4} N$. 
We find that $\Delta t_\textsc{Newton}>\Delta t_\textsc{FSSI}$ for every $N$, and that FSSI is $\sim 6\times 10^{3}$ faster than Newton-Raphson for large $N$. This shows that, in the lack of a good first guess, FSSI can be better than Newton-Raphson method for every value of $N$. Of course, for small $N$ the specific, semi-analytic routine scipy.lambertw \cite{SciPyLambertW}, having $\Delta t_\textsc{scipy\_LambertW}=3.8\times 10^{-7} N$ outperforms the FSSI, but surprisingly the opposite is true in the large $N$ regime, in which FSSI is $\sim20$ times faster than scipy.lambertw.

Finally, for Lambert W with $n=10^4$, corresponding to tolerance $2\times 10^{-13}$, we obtain a combined behavior 
$\Delta t_\textsc{FSSI}\simeq 2.1\times 10^{-8} N + 1.1\times 10^{-3}$ while 
$\Delta t_\textsc{Newton}=1.3\times 10^{-4} N$. We find that $\Delta t_\textsc{Newton}>\Delta t_\textsc{FSSI}$ for every $N\gtrsim 8 $, and that FSSI is $\sim 7\times 10^{3}$ faster than Newton-Raphson for large $N$. This shows that, in the lack of a good first guess, FSSI is much better than Newton-Raphson method for every value of $N$. Again, for large $N$,  $\Delta t_\textsc{FSSI}<\Delta t_\textsc{scipy\_LambertW}=3.8\times 10^{-7} N$ by a factor $\sim20$.

Note that the values of the FSSI execution times are almost equal for Kepler and Lambert problems with the same values of $n$ and $N$. The fact that the method performs at the same speed when applied to functions that are very different from each other is a further proof of its universality.

\section{\label{sec:conclusions}Conclusions}

In this study,  we described a scheme, called FSSI, based on switch and spline to invert monotonic functions under very general conditions.  Moreover,  we derived analytical expressions for the associated  theoretical errors of this method, and tested it on examples that are of interest in physics, including the computation of Lambert W function and the solution of Kepler's equation.
As a summary,  the FSSI   method has several advantages over other more standard techniques for inverting functions:
\begin{itemize}
    \item
    It is simple and universal and, unlike Newton methods, it does not require any initial guess.
    \item
    The error is much smaller than what could be expected from general spline analysis, by a $\sim 10^{-22}$ factor for  $W\in [0,10]$ or by a factor $2\times 10^{-4}$ for Kepler problem.
    \item
    This scheme is superior to, and much faster than, Newton-Raphson method when the latter is difficult to apply, when no good first guess is available, or when the values of the inverse function are required on an entire interval or in a large number of different points.
    \item
   When applied to Kepler's problem (e.g. with eccentricity $\text{e}=0.8$), FSSI becomes faster than Newton's methods for $N$ greater than a few points, and is $\sim 2\times 10^3$ times faster for large $N$. If the requested accuracy is of the order of $10^{-6}$ rad, which is a low enough value for most applications, the speed of the FSSI algorithm is faster than Newton's for $N\ge 2$.
    \item
    The $N$ dependence of the scheme can be described as $\mathcal{O}(1)+\epsilon\, \mathcal{O}(N)$. For a wide range of  $N$ values, the $\mathcal{O}(N)$ behavior is negligible as compared with the $\mathcal{O}(1)$ behavior; however for very large $N$, when the $\mathcal{O}(N)$ part dominates, the linear coefficient $\epsilon$  is several orders of magnitude smaller than those of the other methods available.  

\end{itemize}

For all these reasons, we believe that this method could become a competitive choice  for inverting functions in a wide range of applications, and the first choice for solving Kepler's equation.

\section{\label{sec:acknowledgments}Acknowledgements}

We thank A. Paredes, D. González-Salgado and H. Michinel for discussions. This work has been supported by grants FIS2017-83762-P from Ministerio de Econom\'\i a y Competitividad (Spain), and grant GPC2015/019 from Conseller\'\i a de Cultura, Educaci\'on e Ordenaci\'on Universitaria (Xunta de Galicia).

\end{document}